\documentclass{article}
\usepackage{epsfig}

\DeclareGraphicsExtensions{eps,ps,epsi}
\small\normalsize

\newcommand{\be}{\begin{equation}}
\newcommand{\ee}{\end{equation}}
\newcommand{\bm}[1]{\mbox{\boldmath $#1$}}

\title{Connection between the Dielectric and the Ballistic Treatment of Collisional Absorption}
\author{R. Schneider\thanks{e-mail:Ralf.Schneider@physik.tu-darmstadt.de}\\
Theoretical Quantum Electronics, Institute of Applied Physics,\\ Darmstadt University of Technology,\\
Hochschulstr. 4a}
\date{}

\begin{document}

\maketitle

\begin{abstract}
In this work two important models of treating {\em collisional absorption in a laser driven plasma} are compared, 
the dielectric and the ballistic
model. We will see that there exists a remarkable connection between these basic approaches 
which could give a hint how to overcome the inherent limitations. The approximations made in the models are not 
identical and lead to different advantages and disadvantages.\\
We notice that the dieletric model is able to handle screening in a selfconsistent manner, but is limited to first order
in the electron-ion interaction. The ballistic model calculates the electron-ion collision exactly in each
order of the interaction, but has to introduce a cut-off to incorporate screening effects. This means in the context of kinetic
theory that the electron-ion correlation has to be calculated either in {\em random phase} or in
{\em ladder} approximation, or, in other words, the linearized {\em Lenard-Balescu} or {\em Boltzmann}
collision term has to be used.
\end{abstract}

\section{Basic Results}
\subsection{The Ballistic Model}
\footnote{The model is based on the usage of the Coulomb cross section which is the subject of standard text books. 
For a detailed discussion of the ensemble averaging and the Coulomb logarithm see ref.~\cite{Ref1}.}
The momentum loss per unit time along the initial direction of a electron scattered by an ion reads
\[
 \dot{\bm p} = - m_e \nu_{ei} ({\bm v}) {\bm v} = - \frac{K}{v^3} {\bm v}, \;\;
 K = \frac{Z^2 e^4 n_i}{4 \pi \varepsilon^2_0 m_e} \ln \Lambda, \;\; 
\ln\Lambda=\frac{1}{2}\ln\frac{b^2_{\rm max}+b^2_{\bot}}{b^2_{\rm min}+b^2_{\bot}}\;\;.
\]
This equation defines the collision frequency $\nu_{ei} ({\bm v})$. The Coulomb logarithm $\ln\Lambda$ depends on two cut-off
lengths $b_{\rm max}$ and $b_{\rm min}$ which describe the dynamical screening of the Coulomb potential
and the quantummechanical closing of the singularity at the origin on the scale of a De Broglie wavelength. So we assume
\[
b_{\rm max}=\frac{\sqrt{\hat{v}^2_{\rm os}/2+v^2_{\rm th}}}{\max(\omega,\omega_p)}\;,\;
b_{\rm min}=\frac{\hbar}{m_e\sqrt{\hat{v}^2_{\rm os}/2+v^2_{\rm th}}}\;\;.
\]
Notice that the collision parameter $b_\bot$ for perpendicular deflection is an inherent quantity for the Coulomb collision
and not a cut-off.\\
Calculating the ensemble average over an isotropic distribution function, where the
Coulomb logarithm is treated as a constant, we could determine the 
{\em time-dependent collision frequency}
\[
\nu_{ei} (t) = \frac{K}{m_e v^3_{\rm os} (t)} \int_0^{v_{\rm os}(t)} 4 \pi
v^2_e f (v_e) d v_e\;\;.
\]
In order to compare this result with the dielectric model, we have to determine the time averaged energy absorption 
of the plasma in the laser field for a harmonic electron movement. The energy absorption is connected to the 
{\em time-averaged collision frequency} $\overline{\nu}_{ei}$ by
\[
m_e \overline{\nu_{ei} {\bm v}^2_{\rm {os}}} = m_e \overline{\nu_{ei}} \overline{{\bm v}^2_{
\rm {os}}} = 2 \overline{\nu}_{ei} \overline{E}_{\rm {kin}}\;\;.
\]
Hence, we find for the cycle averaged absorped energy density $\overline{\dot{\cal E}}$
\be\label{Eb_eq}
\overline{\dot{\cal E}}=2n_e\overline{\nu}_{ei}\overline{E}_{\rm {kin}} =
Z\omega^4_p m_e\ln\Lambda\;\overline{\frac{1}{v_{\rm{os}} (t)}\int^{v_{\rm{os}}(t)}_0 v_e^2 f (v_e) d v_e }\;\;.
\ee

\subsection{The Dielectric Model}
In many papers about collisional absorption in plasmas the dielectric theory was the starting point, refs.~\cite{Ref4},~\cite{Ref5},~\cite{Ref6}.
As this theory is well known we only present the result for the {\em cycle averaged absorped energy density}
\be\label{Ed_eq}
\overline{\dot{\cal E}}=\frac{Z\omega_p^4 m_e}{\pi^2 \hat{v}_{\rm os}}\int_0^{k_{\rm max}}\frac{dk}{k}\;\;
F\left(k,\omega,\frac{\hat{v}_{\rm os}}{v_{\rm th}}\right)
\ee
\be\label{F_eq}
F(k,\omega,\frac{\hat{v}_{\rm os}}{v_{\rm th}})=\omega^2\sum_{n=1}^\infty n \;\Im\{\epsilon_n^{-1}\}
\int_0^{\frac{k\hat{v}_{\rm os}}{\omega v_{\rm th}}} dx J^2_n(x)
\ee
\[
\epsilon_n(k,\omega)=1+\frac{1}{k^2}-\sqrt{2}\frac{n\omega}{k^3}D\left(\frac{n\omega}{\sqrt{2}k}\right)-
i\sqrt{\frac{\pi}{2}}\frac{n\omega}{k^3}e^{-\frac{n^2\omega^2}{2 k^2}}
\]
with
\[
D(x)=e^{-x^2}\int_0^x e^{t^2}dt,
\;\;k\rightarrow k/k_D\;,\;k_D=\frac{\omega_p}{v_{\rm th}}\;,\;\omega\rightarrow \omega/\omega_p\;\;.
\]
The upper integral limit $k_{\rm max}$ in eq.~(\ref{Ed_eq}) is necessary in the classical case due to the divergence
of the integral for large $k$. In the quantum case an additional term $\exp(-{k^2}/{8k^2_{\rm B}})$ ($k_{\rm B}$ De Broglie wavenumber)
appears inside the integral of eq.~(\ref{Ed_eq}), which confirms the assumption that the De Broglie wavelength 
has to be considered in $k_{\rm max}$, refs.~\cite{Ref2},~\cite{Ref3}.

\section{The Connection between the Models}
\begin{figure}
{\unitlength1mm
\begin{picture}(70,85)
\put(0,0){
\begin{picture}(70,85)
\put(-5,30){\epsfig{file=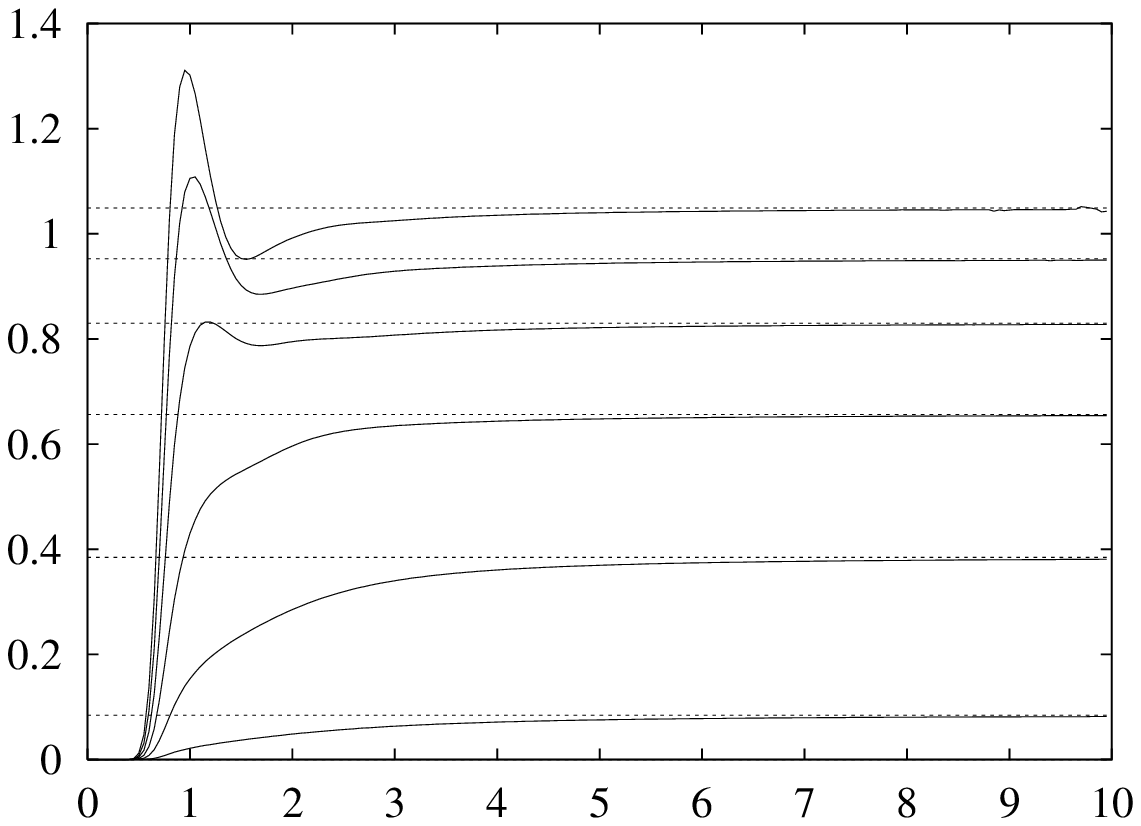, width=7cm}}
\put(27,25){\small $k/k_{\rm D}$}
\put(-2,20){\parbox[t]{7.5cm}{\small\refstepcounter{figure}\label{F_G_fig}Fig.~\ref{F_G_fig}: 
The integral kernel $F(k,\omega,\frac{\hat{v}_{\rm os}}{v_{\rm th}})$ (solid, eq.~(\ref{F_eq})) and 
$G(\frac{\hat{v}_{\rm os}}{v_{\rm th}})$ (dashed, eq.~(\ref{G_eq})) 
for $\omega/\omega_p=2$ and $0\le\hat{v}_{\rm os}/v_{\rm th}\le 6$ (bottom to top).}}
\end{picture}
}
\put(85,0){
\begin{picture}(70,85)
\put(-5,30){\epsfig{file=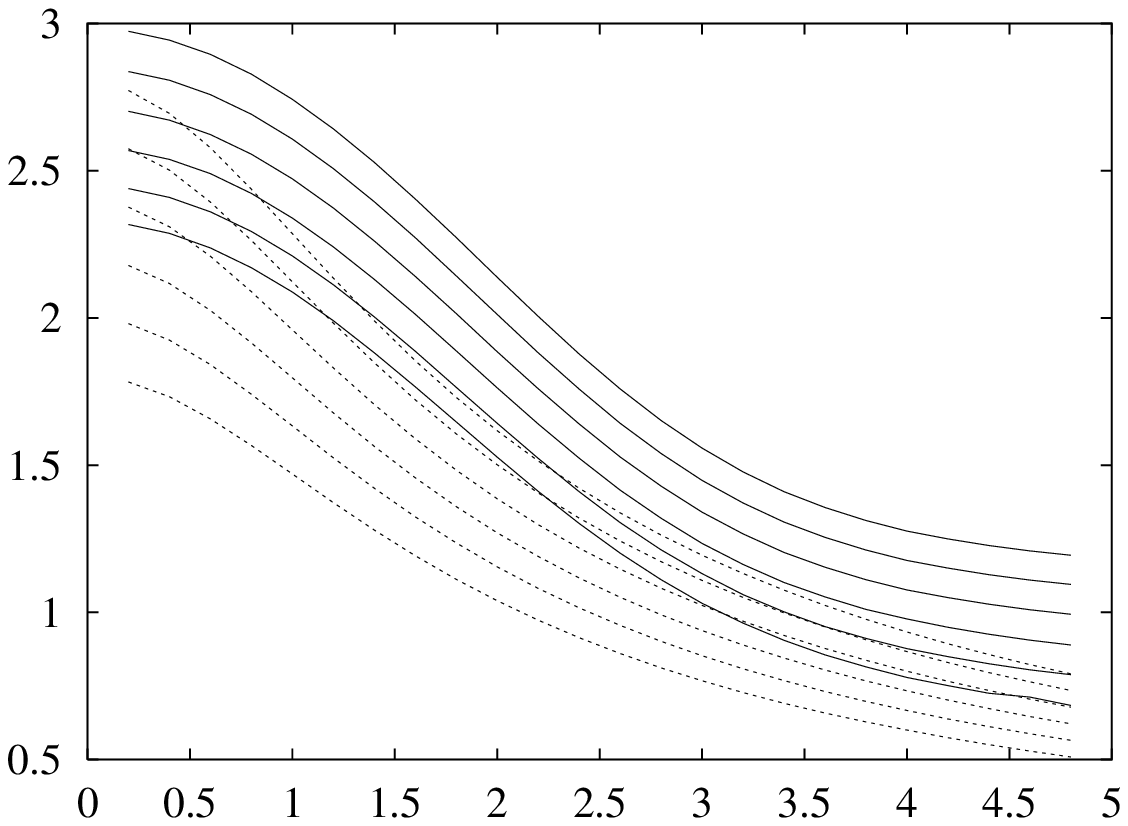, width=7cm}}
\put(27,25){\small $\hat{v}_{\rm os}/v_{\rm th}$}
\put(-2,20){\parbox[t]{7cm}{\small\refstepcounter{figure}\label{kcut_fig}Fig.~\ref{kcut_fig}: 
$k_{\rm min}$ determined by the dielectric model (solid) and the
one assumed in the ballistic model $b^{-1}_{\rm max}$ (dashed) for $1.8\le\omega/\omega_p\le 2.8$ (bottom to top).}}
\end{picture}
}
\end{picture}
}
\end{figure}

When analizing the function $F(k,\omega,\frac{\hat{v}_{\rm os}}{v_{\rm th}})$ we get the remarkable equality
\be\label{G_eq}
\lim_{k\rightarrow\infty}F(k,\omega,\frac{\hat{v}_{\rm os}}{v_{\rm th}})=G(\frac{\hat{v}_{\rm os}}{v_{\rm th}})= 
\pi^2\hat{v}_{\rm os}\overline{\frac{1}{v_{\rm os}(t)}\int_0^{v_{\rm os}(t)} v_e^2f_{\rm M}(v_e)\;dv_e}\;\;,
\ee
which connects eq.~(\ref{Eb_eq}) and eq.~(\ref{Ed_eq}) if $f(v_e)$ is set Maxwellian, see Fig.~\ref{F_G_fig}.\\

The approximation that the $k$-dependence of
$F(k,\omega,\frac{\hat{v}_{\rm os}}{v_{\rm th}})$ is a theta function leads us to the Coulomb logarithm
\be\label{theta_eq}
\int\limits_0^{k_{\rm max}}\frac{dk}{k}\;F(k,\omega,\frac{\hat{v}_{\rm os}}{v_{\rm th}})\approx
G(\frac{\hat{v}_{\rm os}}{v_{\rm th}})\ln\frac{k_{\rm max}}{k_{\rm min}} = 
G(\frac{\hat{v}_{\rm os}}{v_{\rm th}})\ln\frac{b_{\rm max}}{b_{\rm min}}\;\;.
\ee
The lower cut-off $k_{\rm min}$, which is nothing else the inverse screening length, 
will be determined by comparing the integrals
\[
\int_0^{k_0} dk\;F(k,\omega,\frac{\hat{v}_{\rm os}}{v_{\rm th}})=
\int_0^{k_0} dk\;G(\frac{\hat{v}_{\rm os}}{v_{\rm th}})\Theta(k-k_{\rm min}),
\]
where $k_0$ is chosen large enough that $F(k_0,\omega,\frac{\hat{v}_{\rm os}}{v_{\rm th}})$ and 
$G(\frac{\hat{v}_{\rm os}}{v_{\rm th}})$ are equal. \\
Comparing the dielectric inverse screening length $k_{\rm min}$ and
the one introduced in the ballistic model, Fig.~\ref{kcut_fig}, we come to a good qualitative agreement.
Nervertheless, a quantitative difference appears. It must be kept in mind that we handled
the Coulomb logarithm as a constant during the 
ensemble average and also during the time average, which is not done in the dielectric model. The discrepancy should decrease if we overcome
this approximation, which will be the subject of further investigations.

\section{Conclusions}
It was shown in the previous section that there exists a strong connection between the dielectric and the ballistic model.
This results from the fact that the integral kernel $F(k,\omega,\frac{\hat{v}_{\rm os}}{v_{\rm th}})$, eq.~(\ref{F_eq}), 
only becomes a function of $v_{\rm os}/v_{\rm th}$ and agrees 
with the integral term of eq.~(\ref{Eb_eq}). When calculating  it is essential to include enough orders of Bessel functions
for large $k$. So, as the integral in eq.~(\ref{F_eq}) runs up to large $k$, it is never a good approximation to take only
a few orders of Bessel functions, which was done by many authors to get analytical expressions for the absorption.
Furthermore, it is much easier to find approximations of the term in eq.~(\ref{Eb_eq}), ref.~\cite{Ref1}, 
than of the complicated expression eq.~(\ref{F_eq}).\\
When making the approximation eq.~(\ref{theta_eq}) in the dielectric treatment we could see the difference between both models. In case
of the dielectric model the collision parameter $b_\bot$ for perpendicular deflection is missing. This is exactly the term
which leads beside the De Broglie wavelength to the convergence of the collision integral for small collision parameters which
means large $k$ in eq.~(\ref{Ed_eq}). The disappearance of that length is a consequence of the weak coupling approximation in the dielectric
theory, equivalent to the first order Born approximation or straight orbit assumption. We could expect that the integral
kernel $F(k,\omega,\frac{\hat{v}_{\rm os}}{v_{\rm th}})$ should show a decay to zero for $k>b^{-1}_\bot$ when we go beyond 
the weak coupling approximation, which leads to a reduced absorption.
This is in agreement to stopping power calculations, ref.~\cite{Ref7}, where the authors found  
an overestimation of the stopping power in the case of the first order Born approximation in the electron-ion coupling. Including the static
shielded T matrix they found good agreement with numerical results.

\end{document}